\documentclass[11pt]{article}
\usepackage[a4paper,margin=2.35cm]{geometry}
\usepackage{amsmath,amssymb,mathtools,bm}
\usepackage{booktabs}
\usepackage{graphicx}
\usepackage{microtype}
\usepackage{hyperref}
\usepackage{enumitem}
\hypersetup{colorlinks=true,linkcolor=blue,citecolor=blue,urlcolor=blue}
\newcommand{\cH}{\mathcal H}
\newcommand{\cK}{\mathcal K}
\newcommand{\cB}{\mathcal B}
\newcommand{\R}{\mathbb R}
\newcommand{\dd}{\mathrm d}

\title{\bf Bell--CHSH Violation for a Massless Majorana Field\\in a Double Cone from Modular Localization}
\author{S. P. Sorella\\[2mm]\small Instituto de F\'isica Armando Dias Tavares, Universidade do Estado do Rio de Janeiro\\\small Rua S\~ao Francisco Xavier 524, 20550-900 Maracan\~a, Rio de Janeiro, Brazil\\\small \texttt{silvio.sorella@fis.uerj.br}}
\date{}
\begin{document}
\maketitle

\begin{abstract}
We investigate Bell--CHSH correlations for a chiral massless Majorana field localized in a double cone. The conformal modular flow of the associated light-ray interval is converted into translations by a rapidity-like coordinate, and the one-particle scalar product is derived in modular momentum space. Particular attention is paid to the fact that this scalar product carries a Fermi--Dirac weight. Requiring the modular conjugation to be antiunitary with respect to that weighted scalar product fixes the Fourier-space realization of the modular objects. The resulting Tomita--Takesaki operator and its adjoint generate Alice and twisted-dual Bob directions for which all four crossed CAR orthogonality relations follow automatically from modular localization. The Bell problem then reduces to a simple one-function functional. For a Gaussian modular-momentum profile $h_\sigma(k)=\exp[-k^2/(2\sigma^2)]$ the Bell parameter is larger than the classical bound for a broad range of widths and approaches the Tsirelson value analytically,
\[
\lim_{\sigma\to0^+}\cB(\sigma)=2\sqrt2.
\]
Thus near-maximal Bell violation is obtained from an explicit sequence of smooth rapidly decreasing profiles concentrated around zero modular momentum, corresponding to the modular spectral value $\lambda=1$.
\end{abstract}

\section{Introduction}
Bell inequalities provide one of the sharpest distinctions between classical local correlations and quantum theory \cite{Bell,CHSH}. In relativistic quantum field theory, observables are attached to spacetime regions, the vacuum is entangled across causally disjoint regions, and local von Neumann algebras are naturally organized by Tomita--Takesaki modular theory \cite{Takesaki}. The connection between Bell correlations and local quantum field theory was established in a series of works by Summers and Werner \cite{SW1,SW2,SW3}.

For wedge regions modular localization is particularly explicit through the Bisognano--Wichmann property \cite{BW}.  Here we carry out an analogous construction for a bounded conformal region: a massless chiral Majorana field localized in an interval, corresponding to a double cone in two-dimensional Minkowski spacetime.  Unlike a wedge, a double cone is bounded not only in space but also in time; it therefore provides a more localized setting for finite-region correlations.  The modular group preserving the interval is a M\"obius flow, and a change of variable converts it into a translation, making the modular spectral variable explicit \cite{BGL93,BGL02,Guido}.

Exact modular analyses of free fermions in multicomponent regions have been developed by Casini and Huerta \cite{CasiniHuerta}.  Related calculations for chiral theories and two-interval geometries further clarify the fermionic modular structure \cite{AriasEtAl}.  Explicit modular conjugations for massless Dirac fields in multicomponent regions were obtained in Refs.~\cite{AbateEtAl,MintchevTonni}.

The bounded spacetime region considered below is the double cone
\begin{equation}\label{eq:double-cone}
D_R=\{(t,x)\in\R^{1,1}: |x|+|t|<R\}.
\end{equation}
At the abstract algebraic level, Summers and Werner proved maximal Bell correlations for tangent spacetime regions, including tangent double cones \cite{SW3}.  A complementary explicit route to near-Tsirelson violation for free spinor fields using Carleman and Hankel operators has recently been given in Ref.~\cite{DudalVandermeersch}.  The purpose of the present construction is concrete: we use the modular description of a bounded interval to produce explicit one-particle directions, localized Majorana observables, and a computable Bell functional.

The construction has four steps.  We derive the one-particle scalar product of the chiral Majorana field in the interval and its Fermi--Dirac weight in modular momentum, reconstruct the modular operators with all adjoints taken in that weighted Hilbert space, generate explicit Alice directions from the Tomita--Takesaki operator and Bob directions from its twisted dual, and finally form normalized local dichotomic observables.  The correction is conceptually important: once the weighted scalar product is used consistently, the crossed real parts required by the CAR vanish automatically.  In the resulting Bell test Alice's observables are localized in $D_R$, whereas Bob's observables are localized in the causal complement, with the fermionic relation between the two local algebras expressed by twisted Haag duality.

The bounded-region setting is more restrictive than the wedge construction: its modular flow is M\"obius rather than a geometric boost, and the local directions must be built within the standard subspace of a finite interval.  The modular coordinate makes this problem tractable by converting the flow into translations, while twisted duality identifies the complementary Bob directions.  The resulting Bell functional is elementary, and a sequence of smooth Gaussian modular-momentum profiles approaches $2\sqrt2$, so that the Tsirelson value is the supremum of an explicit regular sequence.

\section{Massless Majorana field in a double cone}
Starting from the double cone in Eq.~\eqref{eq:double-cone}, introduce light-ray coordinates
\begin{equation}
u=t+x,\qquad v=t-x.
\end{equation}
Then $D_R$ is the square $-R<u<R$, $-R<v<R$. For one chiral component only one null coordinate is relevant; choosing $u$, localization is characterized by the interval
\begin{equation}
I_R=(-R,R).
\end{equation}
The chiral Majorana field on the light ray is represented, in a convenient
normalization, by
\begin{equation}
\psi(u)=\int_0^\infty \frac{\dd p}{\sqrt{2\pi}}\left(a_p e^{-ipu}+a_p^\dagger e^{ipu}\right),
\qquad \{a_p,a_q^\dagger\}=\delta(p-q).
\end{equation}
For a real test function $f\in C_0^\infty(I_R)$,
\begin{equation}
\psi(f)=\int_{-R}^{R}\dd u\,\psi(u)f(u).
\end{equation}
The vacuum two-point function induces the one-particle scalar product whose
completion, after complexification, defines the one-particle Hilbert space
$\cH_1$:
\begin{equation}
\langle f|g\rangle
=\langle0|\psi(f)\psi(g)|0\rangle
=\frac{1}{2\pi i}\int \dd u\,\dd u'\,
\frac{f(u)g(u')}{u-u'-i0},
\end{equation}
for real smearings. The real closure of the localized vectors is a standard real subspace $\cK(D_R)$ \cite{BGL93,BGL02,Guido}.  Explicitly, standardness means
\begin{equation}\label{eq:standardness}
\cK(D_R)+i\cK(D_R)=\cH_1,
\qquad
\cK(D_R)\cap i\cK(D_R)=\{0\}.
\end{equation}

\subsection{Modular coordinate}
Introduce the rapidity-like coordinate
\begin{equation}
u=R\tanh\rho,\qquad
\rho=\frac12\log\frac{R+u}{R-u}.
\end{equation}
Thus, as $u$ ranges bijectively over the interval $(-R,R)$, the modular coordinate $\rho$ ranges bijectively over the whole real line.
The one-parameter M\"obius transformation preserving $I_R$ is
\begin{equation}
u_t=R\,\frac{u+R\tanh(\pi t)}{R+u\tanh(\pi t)}.
\end{equation}
In the modular coordinate it becomes the translation
\begin{equation}
\rho\longmapsto \rho+\pi t.
\end{equation}
Thus the generator of the modular flow is $K=-i\partial_\rho$, and on
test functions
\begin{equation}\label{eq:modular-flow-rho}
(\delta^{it}f)(\rho)=f(\rho-\pi t),
\qquad
\delta^{it}=e^{-i\pi tK}.
\end{equation}
This is the double-cone conformal analogue of the
Bisognano--Wichmann description for wedges: the M\"obius modular action
preserving a bounded interval is linearized into translations in the
coordinate $\rho$.
With the conformal Jacobian appropriate to a chiral field of dimension $1/2$ absorbed into the test function\footnote{More explicitly, if $F(u)$ is the original light-ray smearing and $\psi_u$ is the chiral Majorana field, write $\psi_\rho(\rho)=(du/d\rho)^{1/2}\psi_u(u(\rho))$.  Then $\int du\,F(u)\psi_u(u)=\int d\rho\,f(\rho)\psi_\rho(\rho)$, where $f(\rho)=(du/d\rho)^{1/2}F(u(\rho))=\sqrt R\,\operatorname{sech}\rho\,F(R\tanh\rho)$.  This redefinition is the Jacobian absorbed in the Majorana test function.}, the scalar product becomes
\begin{equation}
\langle f|g\rangle
=\frac{1}{2\pi i}\int_{\R^2}\dd\rho\,\dd\rho'\,
\overline{f(\rho)}\,
\frac{1}{\sinh(\rho-\rho'-i0)}\,g(\rho'),
\end{equation}
where the displayed sesquilinear form denotes the complexified one-particle product.

\section{Fourier-space scalar product and modular data}
We use the Fourier convention
\begin{equation}
f(\rho)=\frac1{\sqrt{2\pi}}\int_\R \dd k\,e^{-ik\rho}\widehat f(k),
\qquad
\widehat f(k)=\frac1{\sqrt{2\pi}}\int_\R \dd\rho\,e^{ik\rho}f(\rho).
\end{equation}
For real Majorana test functions, substitution in the scalar product above,
followed by the variables $x=\rho-\rho'$ and $y=\rho'$, makes the $y$
integration produce $\delta(k+k')$.  Thus
\begin{equation}\label{eq:scalar-reflected}
\langle f|g\rangle=\int_\R\dd k\,
\widehat f(k)\widehat g(-k)\,\mu(k),
\end{equation}
where
\begin{equation}\label{eq:mu-def}
\mu(k)=\frac{1}{2\pi i}\int_\R\dd x\,
\frac{e^{-ikx}}{\sinh(x-i0)}.
\end{equation}
We evaluate this Fourier transform by the Cauchy residue theorem.  The poles
of $1/\sinh z$ are separated by $i\pi$ and their residues alternate in sign.
For $k>0$, closing the contour in the lower half-plane gives
\begin{align}
\mu(k)
&=-\sum_{n=1}^{\infty}(-1)^n e^{-n\pi k}
=e^{-\pi k}\sum_{m=0}^{\infty}(-e^{-\pi k})^m \notag\\
&=\frac{e^{-\pi k}}{1+e^{-\pi k}}
=\frac{1}{1+e^{\pi k}},
\qquad k>0.
\label{eq:mu-positive}
\end{align}
For $k<0$, one closes in the upper half-plane and obtains instead
\begin{align}
\mu(k)
&=\sum_{n=0}^{\infty}(-1)^n e^{n\pi k}
=\frac{1}{1+e^{\pi k}},
\qquad k<0.
\label{eq:mu-negative}
\end{align}
Hence, for every real $k$,
\begin{equation}\label{eq:fermi-dirac-kernel}
\boxed{\displaystyle \mu(k)=\frac{1}{1+e^{\pi k}}.}
\end{equation}
Inserting this result in Eq.~\eqref{eq:scalar-reflected} yields the
intermediate expression
\begin{equation}\label{eq:scalar-intermediate}
\langle f|g\rangle=\int_\R\dd k\,
\frac{\widehat f(k)\widehat g(-k)}{1+e^{\pi k}}.
\end{equation}
Finally, changing $k\mapsto-k$ and using the Majorana reality relation
$\widehat f(-k)=\overline{\widehat f(k)}$ gives the
one-particle inner product
\begin{equation}\label{eq:inner}
\boxed{
\langle f|g\rangle
=\int_\R \dd k\,w(k)\,\overline{\widehat f(k)}\widehat g(k),
\qquad
w(k)=\frac1{1+e^{-\pi k}}.}
\end{equation}
The crucial identity is
\begin{equation}\label{eq:weightid}
w(-k)=e^{-\pi k}w(k),\qquad w(k)+w(-k)=1.
\end{equation}

\subsection{Weighted modular conjugation}
The modular flow has already been identified in Eq.~\eqref{eq:modular-flow-rho}.
Its analytic continuation to $t=-i/2$ acts in momentum space as
\begin{equation}\label{eq:delta}
(\delta^{1/2}f)(k)=e^{\pi k/2}f(k).
\end{equation}
Because the inner product is weighted, the reflection--conjugation map must carry the corresponding Radon--Nikodym factor. Requiring $j$ to be antiunitary and involutive gives
\begin{equation}\label{eq:j}
\boxed{(jf)(k)=e^{-\pi k/2}\overline{f(-k)}.}
\end{equation}
Indeed, using Eq.~\eqref{eq:weightid},
\begin{equation}
\|jf\|^2=\int \dd k\,w(k)e^{-\pi k}|f(-k)|^2
=\int \dd k\,w(k)|f(k)|^2.
\end{equation}
Moreover
\begin{equation}
j^2=1,\qquad j\delta j=\delta^{-1}.
\end{equation}
The Tomita--Takesaki operator is therefore
\begin{equation}\label{eq:s}
\boxed{(sf)(k)=(j\delta^{1/2}f)(k)
=e^{-\pi k}\overline{f(-k)}.}
\end{equation}
It obeys $s^2=1$ on its natural domain. Its antilinear adjoint must be computed with the same weighted scalar product. A direct calculation gives
\begin{equation}\label{eq:sadj}
\boxed{(s^\dagger f)(k)=\overline{f(-k)}= (j\delta^{-1/2}f)(k).}
\end{equation}
Thus
\begin{equation}
(s^\dagger)^2=1.
\end{equation}
Equations~\eqref{eq:delta}--\eqref{eq:sadj} are the Fourier-space modular data used below.  Since $\delta^{1/2}$ is the unbounded multiplication operator by $e^{\pi k/2}$, it is useful to give explicitly its domain, which also specifies the natural domain of the Tomita--Takesaki operator $s=j\delta^{1/2}$:
\begin{equation}
\mathrm{Dom}\,\delta^{1/2}
=\left\{f\in\cH_1:\int_\R \dd k\,
\frac{e^{\pi k}|f(k)|^2}{1+e^{-\pi k}}<\infty\right\}.
\end{equation}
All Gaussian profiles used below belong to a common invariant core for the modular operators.

\section{Tomita--Takesaki localization of the Bell directions}
At one-particle level
\begin{equation}
\cK(D_R)=\{\xi\in\mathrm{Dom}\,s:s\xi=\xi\}.
\end{equation}
Let $h(k)$ be real, smooth, even and rapidly decreasing. Define Alice's two directions by
\begin{equation}\label{eq:alice}
f=(1+s)h,\qquad f'=(1+s)(ih).
\end{equation}
Using Eq.~\eqref{eq:s},
\begin{align}
\boxed{f(k)=(1+e^{-\pi k})h(k),}\label{eq:f}\\
\boxed{f'(k)=i(1-e^{-\pi k})h(k).}\label{eq:fp}
\end{align}
Both satisfy $sf=f$ and $sf'=f'$.

The symplectic dual is characterized by $s^\dagger q=q$. Put
\begin{equation}
t(k)=\tanh\frac{\pi k}{2}
\end{equation}
and choose
\begin{equation}\label{eq:q12}
\boxed{q_1(k)=h(k),\qquad q_2(k)=i\,t(k)h(k).}
\end{equation}
Since $h$ is real even and $t$ real odd,
\begin{equation}
s^\dagger q_1=q_1,\qquad s^\dagger q_2=q_2.
\end{equation}
Thus $q_1,q_2\in\cK(D_R)'$. At the local-algebra level Bob's observables are obtained from these dual directions by the usual fermionic twist/Klein transformation, the one-particle counterpart of twisted Haag duality \cite{CasiniHuerta,AriasEtAl,AbateEtAl,MintchevTonni,CaribeEtAl}. We keep the twist explicit in the Bell correlator below; equivalently one may denote the twisted directions by $g_j=iq_j$.

\subsection{Automatic crossed CAR orthogonality}
For real Majorana smearings,
\begin{equation}
\{\psi(u),\psi(v)\}=2\,\mathrm{Re}\langle u|v\rangle\,\mathbf1.
\end{equation}
Define
\begin{equation}
H[h]=\int_\R\dd k\,h(k)^2,
\qquad
T[h]=\int_\R\dd k\,t(k)^2h(k)^2.
\end{equation}
Using Eqs.~\eqref{eq:inner}, \eqref{eq:f}--\eqref{eq:q12}, one obtains
\begin{align}
\langle f|q_1\rangle &=H, & \langle f|q_2\rangle&=0,\\
\langle f'|q_1\rangle&=0, & \langle f'|q_2\rangle&=T.
\end{align}
After the fermionic twist $q_j\mapsto iq_j$, all four Alice--Bob pairings are purely imaginary or zero. Hence
\begin{equation}\label{eq:crossed}
\boxed{
\mathrm{Re}\langle f|iq_1\rangle=
\mathrm{Re}\langle f|iq_2\rangle=
\mathrm{Re}\langle f'|iq_1\rangle=
\mathrm{Re}\langle f'|iq_2\rangle=0.}
\end{equation}
No additional crossed variational constraints are required.

\section{Local dichotomic Majorana observables and the Bell functional}
For every real one-particle direction $f$, the corresponding localized
Majorana observable is defined by
\begin{equation}\label{eq:majorana-observable}
 A(f)=\frac{\psi(f)}{\|f\|}.
\end{equation}
The canonical anticommutation relations imply
\begin{equation}\label{eq:majorana-dichotomic}
 A(f)^\dagger=A(f),\qquad A(f)^2=1,
\end{equation}
so that $A(f)$ is a Hermitian dichotomic observable.  Its vacuum
two-point function is the normalized one-particle scalar product,
\begin{equation}\label{eq:majorana-correlation}
 \langle0|A(f)A(g)|0\rangle=
 \frac{\langle f|g\rangle}{\|f\|\,\|g\|}.
\end{equation}
We now apply these definitions to the localized Alice directions and to
the twisted-dual Bob directions constructed above.  The Alice norms are
\begin{align}
D_+[h]&\equiv\|f\|^2
=\int_\R\dd k\,[1+\cosh(\pi k)]h(k)^2,\label{eq:Dp}\\
D_-[h]&\equiv\|f'\|^2
=\int_\R\dd k\,[\cosh(\pi k)-1]h(k)^2.\label{eq:Dm}
\end{align}
For the dual directions,
\begin{equation}
\|q_1\|^2=\frac{H}{2},\qquad
\|q_2\|^2=\frac{T}{2},
\qquad
\mathrm{Re}\langle q_1|q_2\rangle=0.
\end{equation}
Normalize
\begin{equation}
F=\frac{f}{\sqrt{D_+}},\quad
F'=\frac{f'}{\sqrt{D_-}},\quad
Q_1=\frac{q_1}{\sqrt{H/2}},\quad
Q_2=\frac{q_2}{\sqrt{T/2}}.
\end{equation}
The only nonzero reduced pairings are
\begin{equation}\label{eq:c12}
\boxed{
c_1=\langle F|Q_1\rangle=\sqrt{\frac{2H}{D_+}},
\qquad
c_2=\langle F'|Q_2\rangle=\sqrt{\frac{2T}{D_-}}.}
\end{equation}
Choose the standard $45^\circ$ Bob rotation
\begin{equation}
Q_+=\frac{Q_1+Q_2}{\sqrt2},\qquad
Q_-=\frac{Q_1-Q_2}{\sqrt2}.
\end{equation}
The twisted local Bob observables associated with $Q_\pm$ are Hermitian dichotomic operators. In the one-particle notation the twist contributes the same factor $-i$ as in the standard fermionic Bell construction. Therefore the four vacuum correlators are
\begin{align}
E(F,Q_+)&=\frac{c_1}{\sqrt2}, & E(F,Q_-)&=\frac{c_1}{\sqrt2},\\
E(F',Q_+)&=\frac{c_2}{\sqrt2}, & E(F',Q_-)&=-\frac{c_2}{\sqrt2}.
\end{align}
Let $A=A(F)$ and $A'=A(F')$ denote Alice's normalized Majorana
observables, and let $B=B(Q_+)$ and $B'=B(Q_-)$ denote the corresponding
twisted local Bob observables.  The Bell--CHSH operator is
\begin{equation}\label{eq:CHSHoperator}
\mathcal C=-i\left(AB+A'B+AB'-A'B'\right).
\end{equation}
The four observables entering Eq.~\eqref{eq:CHSHoperator} obey the
standard Bell algebra.  In particular, they are Hermitian and dichotomic,
\begin{equation}
A^\dagger=A,\quad {A'}^\dagger=A',\quad B^\dagger=B,\quad {B'}^\dagger=B',
\qquad
A^2={A'}^2=B^2={B'}^2=1,
\end{equation}
and Alice's observables commute with Bob's observables,
\begin{equation}\label{eq:Bellalgebra}
[A,B]=[A,B']=[A',B]=[A',B']=0.
\end{equation}
For the fermionic theory, this commuting Alice--Bob realization is obtained
after the twist has been implemented in the Bob observables.
The Bell expectation is consequently
\begin{equation}\label{eq:Bellfunctional}
\boxed{
\cB[h]=\langle0|\mathcal C|0\rangle
=\sqrt2\left(c_1+c_2\right)
=\sqrt2\left[
\sqrt{\frac{2H[h]}{D_+[h]}}
+\sqrt{\frac{2T[h]}{D_-[h]}}
\right].}
\end{equation}
This is the corrected one-function Bell functional.

With the Bell algebra above, the standard CHSH identity gives
\begin{equation}
-\mathcal C^2=4-[A,A'][B,B']
\end{equation}
and hence
\begin{equation}
\|\mathcal C\|\le2\sqrt2.
\end{equation}
Therefore the functional \eqref{eq:Bellfunctional} cannot exceed the
Tsirelson value.

\section{Gaussian profiles: numerical violation}
Consider the smooth even profile
\begin{equation}\label{eq:gaussian}
\boxed{h_\sigma(k)=\exp\left(-\frac{k^2}{2\sigma^2}\right),\qquad \sigma>0.}
\end{equation}
All modular-domain requirements are satisfied for every fixed $\sigma>0$. The elementary Gaussian integrals give
\begin{equation}
H_\sigma=\sqrt\pi\,\sigma
\end{equation}
and
\begin{equation}\label{eq:Dgauss}
\boxed{
D_\pm(\sigma)=H_\sigma\left(e^{\pi^2\sigma^2/4}\pm1\right).}
\end{equation}
The remaining integral is
\begin{equation}\label{eq:Tgauss}
T_\sigma=\int_\R\dd k\,
e^{-k^2/\sigma^2}\tanh^2\frac{\pi k}{2}.
\end{equation}
Thus
\begin{align}
c_1(\sigma)&=\sqrt{\frac{2}{1+e^{\pi^2\sigma^2/4}}},\\
c_2(\sigma)&=\sqrt{\frac{2T_\sigma}{H_\sigma\left(e^{\pi^2\sigma^2/4}-1\right)}},\\
\cB(\sigma)&=\sqrt2\,[c_1(\sigma)+c_2(\sigma)].
\end{align}
Direct numerical quadrature of Eq.~\eqref{eq:Tgauss} gives Table~\ref{tab:num}.

\begin{table}[ht]
\centering
\caption{Bell parameter for the Gaussian family.}
\label{tab:num}
\begin{tabular}{rrrr}
\toprule
$\sigma$ & $c_1$ & $c_2$ & $\cB(\sigma)$\\
\midrule
0.500 & 0.837255 & 0.686536 & 2.154967\\
0.300 & 0.943091 & 0.860189 & 2.550223\\
0.200 & 0.975034 & 0.932062 & 2.697042\\
0.100 & 0.993813 & 0.981916 & 2.794102\\
0.050 & 0.998457 & 0.995401 & 2.819740\\
0.020 & 0.999753 & 0.999260 & 2.827032\\
0.010 & 0.999938 & 0.999815 & 2.828078\\
0.005 & 0.999985 & 0.999954 & 2.828340\\
\bottomrule
\end{tabular}
\end{table}

Already at $\sigma=0.5$ the classical bound is violated. As the profile becomes increasingly concentrated near zero modular momentum, the Bell value rapidly approaches $2\sqrt2=2.828427\ldots$.

\begin{figure}[ht]
\centering
\includegraphics[width=.70\linewidth]{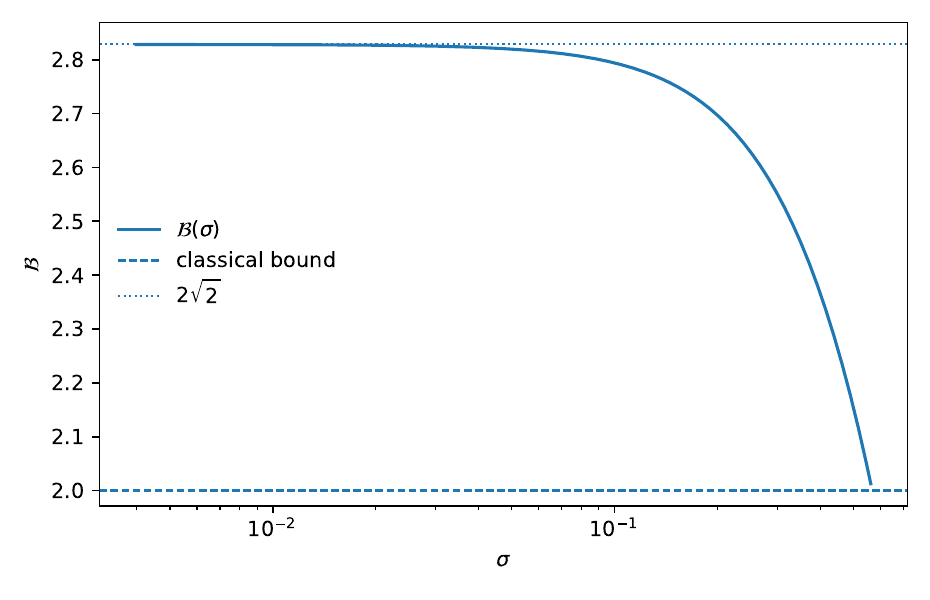}
\caption{The Gaussian Bell parameter $\cB(\sigma)$. The dashed line is the classical CHSH bound $2$ and the dotted line is the Tsirelson value $2\sqrt2$.}
\end{figure}

\section{Analytic approach to the Tsirelson value}
The near-maximal behavior can be proved without numerical optimization. For $k$ near zero,
\begin{align}
\tanh\frac{\pi k}{2}&=\frac{\pi k}{2}+O(k^3),\\
\cosh(\pi k)-1&=\frac{\pi^2k^2}{2}+O(k^4),\\
1+\cosh(\pi k)&=2+\frac{\pi^2k^2}{2}+O(k^4).
\end{align}
For the Gaussian sequence $h_\sigma$, whose normalized mass is concentrated at $|k|=O(\sigma)$, these imply
\begin{equation}
D_+(\sigma)=2H_\sigma\,[1+O(\sigma^2)],
\end{equation}
while
\begin{equation}
D_-(\sigma)=2T_\sigma\,[1+O(\sigma^2)].
\end{equation}
Consequently
\begin{equation}
c_1(\sigma)\longrightarrow1,
\qquad
c_2(\sigma)\longrightarrow1,
\end{equation}
and therefore
\begin{equation}\label{eq:limit}
\boxed{
\lim_{\sigma\to0^+}\cB(\sigma)=2\sqrt2.}
\end{equation}
A more detailed expansion follows from the Gaussian moments. Since
\begin{equation}
\frac{T_\sigma}{H_\sigma}
=\frac{\pi^2\sigma^2}{8}-\frac{\pi^4\sigma^4}{32}+O(\sigma^6),
\end{equation}
we obtain
\begin{align}
c_1(\sigma)&=1-\frac{\pi^2}{16}\sigma^2+O(\sigma^4),\\
c_2(\sigma)&=1-\frac{3\pi^2}{16}\sigma^2+O(\sigma^4),
\end{align}
and hence
\begin{equation}\label{eq:asymp}
\boxed{
\cB(\sigma)=2\sqrt2\left[1-\frac{\pi^2}{8}\sigma^2+O(\sigma^4)\right].}
\end{equation}
Thus the Tsirelson value is the supremum of an explicit sequence of regular Gaussian modular-momentum profiles. No singular profile is required at any finite step; the limit is reached only as the Gaussian width tends to zero.

\section{Discussion and conclusion}
We have presented an explicit finite-region Bell--CHSH construction for a massless chiral Majorana field in a double cone. The modular coordinate converts the M\"obius flow of the interval into translations, while the Fourier representation produces a Fermi--Dirac weight in the one-particle scalar product. The main technical point is that all adjoints and antiunitarity conditions must be taken with respect to this weighted product.

With this requirement, the Fourier-space modular conjugation contains the factor $e^{-\pi k/2}$, the Tomita--Takesaki operator acts as $f(k)\mapsto e^{-\pi k}\overline{f(-k)}$, and its adjoint as $f(k)\mapsto\overline{f(-k)}$. Alice's directions are generated by $1+s$ and the dual directions by the fixed-point space of $s^\dagger$. The fermionic twist then supplies Bob's local observables in the causal complement. The four crossed real parts vanish automatically, as expected from the standard-subspace relation between $\cK(D_R)$ and its symplectic dual.

The normalized Majorana smearings yield local Hermitian dichotomic observables; after the fermionic twist is implemented, the Alice and Bob observables realize the commuting Bell algebra. The resulting Bell functional, Eq.~\eqref{eq:Bellfunctional}, depends on only three elementary quadratic functionals of one real even profile. A Gaussian profile already violates the classical bound for moderate width, and the narrow-Gaussian sequence approaches $2\sqrt2$ both numerically and analytically.

The result also clarifies the role of zero modular momentum. Near $k=0$, the two Alice channels become asymptotically matched to the two dual Bob channels: $D_+\sim2H$ and $D_-\sim2T$. This simultaneous matching is precisely what drives $c_1,c_2\to1$ and hence the Tsirelson limit.  Moreover, Eq.~\eqref{eq:modular-flow-rho} implies $\delta=e^{-\pi K}$, so that modular momentum $k$ corresponds to the modular spectral value $\lambda(k)=e^{-\pi k}$.  Thus $k=0$ corresponds to $\lambda=1$.  In the type $\mathrm{III}_1$ situation of local quantum field theory, this is a point of the continuous modular spectrum, rather than an isolated normalizable eigenvalue; the concentration of the near-maximal violation around $k=0$ therefore probes the distinguished unit point within that continuous spectrum.

Several extensions are natural. One is the geometry of two tangent double cones. For two equal tangent regions related by a spacetime translation, the positive-energy one-particle unitary transports one standard subspace into the other, and the corresponding modular objects are related by unitary conjugation. Another direction is the bosonic counterpart, where bounded self-adjoint observables can be built from superpositions of Weyl operators, for example through odd harmonics or Fej\'er approximants \cite{CaribeEtAlFejer}.

\paragraph{Data availability.}
No experimental or observational data were used. The numerical values in Table~\ref{tab:num} were obtained by direct quadrature of Eq.~\eqref{eq:Tgauss} and the analytic formulas displayed above.

\paragraph{Acknowledgments.}
The author acknowledges financial support from CNPq and is a CNPq researcher under contract 302991/2024-7. The author also acknowledges the use of OpenAI's ChatGPT Plus as an AI-assisted research and writing tool during the preparation of this work. It was used for discussions of intermediate calculations, consistency checks, numerical implementation, and assistance with organization and linguistic revision. All scientific arguments, results, interpretations, and conclusions were independently assessed and remain the sole responsibility of the author.

\end{document}